\documentclass[conference]{IEEEtran}

\usepackage[dvips]{color}
\usepackage{epsf}
\usepackage{times}
\usepackage{epsfig}
\usepackage{graphicx}

\usepackage{amsmath}
\usepackage{amssymb}
\usepackage{amsxtra}
\usepackage{amsthm}
\usepackage{bbm}

\usepackage{here}
\usepackage{rawfonts}
\usepackage{times}
\usepackage{url}
\usepackage{cite}
\usepackage{comment}
\usepackage[utf8]{inputenc}
\usepackage{caption}
\usepackage{subcaption}

\usepackage{pstricks}
\usepackage{algorithm}
\usepackage{algpseudocode}
\usepackage{lipsum}

\makeatletter

\IEEEoverridecommandlockouts 

\begin{document}
\vspace{-0.2cm}
\title{\huge
	Distributional Reinforcement Learning for mmWave Communications with  Intelligent Reflectors on a UAV
}\vspace{-0.2cm}	
\author{ 
	\IEEEauthorblockN{Qianqian Zhang$^1$,  
		Walid Saad$^1$, and Mehdi Bennis$^2$
	} 
	
	\IEEEauthorblockA{\small
		$^1$
		Bradley Department of Electrical and Computer Engineering, Virginia Tech, VA, USA,
		Emails: \url{{qqz93,walids}@vt.edu}. \\
		$^2$Centre for Wireless Communications, University of Oulu, Finland, Email: \url{mehdi.bennis@oulu.fi}. 
		\thanks{This research was supported by the U.S. National Science Foundation under Grant IIS-1633363.}\vspace{-0.6cm}				
	}
}
\maketitle

\setlength{\columnsep}{0.55cm}

\begin{abstract}
In this paper, a novel communication framework that uses an unmanned aerial vehicle (UAV)-carried intelligent reflector (IR)  is proposed to enhance multi-user downlink transmissions over millimeter wave (mmWave) frequencies.   
In order to maximize the downlink sum-rate, the optimal precoding matrix (at the base station) and reflection coefficient (at the IR) are jointly derived.  
Next,  to address the uncertainty of mmWave channels and maintain line-of-sight links in a real-time manner, a distributional reinforcement learning approach, based on quantile regression optimization, is proposed to learn the  propagation environment of mmWave communications, and, then, optimize the location of the UAV-IR so as to maximize the long-term downlink communication capacity. 
Simulation results show that the proposed learning-based deployment of the UAV-IR yields a significant advantage, compared to a non-learning UAV-IR, a static IR, and a direct transmission schemes, in terms of the average data rate and the achievable line-of-sight probability of downlink mmWave communications.

\end{abstract}
 
\IEEEpeerreviewmaketitle

\section{Introduction}

Millimeter wave (mmWave) frequencies are an essential component of next-generation cellular systems, in order to meet the exponential increase of data demand and support a growing number of wireless devices \cite{saad2019vision}. 
Due to a large available bandwidth, mmWave frequencies have a strong potential to provide high communication rates.  
However,  the small wavelength of mmWave spectrum yields the high susceptibility to blockage caused by common objects, such as buildings and foliage, which seriously attenuate mmWave propagation \cite{semiari2017joint}. 

In order to bypass obstacles and prolong the communication range, signal reflectors have been considered as an energy-efficient solution for mmWave communications. 
Signal reflectors can  establish line-of-sight (LOS) links in face of blockage, by replacing a non-line-of-sight (NLOS) mmWave channel by multiple connected LOS links.   
As a passive element, a signal reflector costs no energy and incurs no  additional receiving noise \cite{zhang2019reflections}.   
A reflector-aided transmission enables the connected LOS links  to share the same frequency band, and, thus, yields a high spectrum efficiency.    
Meanwhile, by aligning a large number of low-cost reflective components and jointly inducing phase shifts to incident signals,  an intelligent reflector (IR) can realize beamforming with very lower energy cost for mmWave communications \cite{soorki2019ultra}.   
Moreover, to maintain LOS links in a mobile scenario, an IR can be equipped onto an unmanned aerial vehicle (UAV) platform \cite{zhang2019reflections}, so that the location of the IR can be adjusted intelligently, based on the real-time  communication environment, in order to improve the reliability of mmWave transmissions.  
Compared with a UAV-aided relay station, a UAV-carried IR (UAV-IR) has a simpler antenna structure and smaller power cost.
Therefore, an IR facilitate multiple-input-multiple-output transmissions for the UAV platform, which has very limited onboard energy.    

The use of IRs to improve the communication performance of mmWave has been studied in \cite{zhang2019reflections, soorki2019ultra,  wang2019intelligent, yue2020analysis, cao2019intelligent}.    
In  \cite{zhang2019reflections}, we studied the problem of using a UAV-IR to maximize downlink transmissions towards a single mobile user.    
The authors in \cite{soorki2019ultra} investigated the IR-aided  mmWave communications,  with a deep reinforcement learning (RL) framework, to improve the coverage and increase the data rate in an indoor network.  
In \cite{wang2019intelligent} and \cite{yue2020analysis}, the authors proposed a hybrid precoding scheme to  jointly design the transmit precoding at the base station (BS) and the reflection parameters at the IR.   
However, the prior works in \cite{ zhang2019reflections, soorki2019ultra, wang2019intelligent, yue2020analysis} focus on the downlink transmission to a single user, and they do not consider the more challenging problem of multi-user communications.  
Meanwhile, the works in \cite{cao2019intelligent}  and \cite{guo2019weighted} developed a static IR-assisted mmWave communication framework for multiple downlink users, such that  the weighted sum-rate is maximized.  
However, most of the previous works in \cite{soorki2019ultra,wang2019intelligent,cao2019intelligent, yue2020analysis,guo2019weighted} optimize the reflection transmission of the IR while assuming a fixed location. 
Although a static IR establishes LOS links between transmitters and receivers in the face of blockage, maintaining LOS channels is  challenging, especially in a mobile communication scenario, where the movement of each user dynamically changes the channel state. 
In particular, given the high susceptibility of mmWave signals to human body, a single body rotation can block the LOS link and render the reflection transmission inefficient, using a static IR.  

Machine learning techniques were proposed in in \cite{soorki2019ultra} and  \cite{zhang2020millimeter,taha2019deep,huang2020reconfigurable} in order to address the  wireless channel dynamics and improve the performance of IR-aided communications.  
For instance, in \cite{zhang2020millimeter}, we studied the problem of optimizing beamforming transmissions and reconfigurable reflection of an IR to serve multiple users, using a distributional RL method, so as to maximize the  downlink sum-rate.    
The authors in \cite{taha2019deep} proposed a deep learning approach to optimize the IR phase shift, given a limited number of active reflective elements.
The work in \cite{huang2020reconfigurable} introduced an  RL-based scheme to jointly optimize the precoding transmission and signal reflection.    
However, all the prior art in \cite{soorki2019ultra} and \cite{zhang2020millimeter,taha2019deep,huang2020reconfigurable} does not consider a mobile IR whose location can be dynamically optimized so as to enhance the mmWave reflection performance.

The main contribution of this paper is, thus, a novel  downlink framework using   a UAV-IR to assist a mmWave BS for multi-user  communications. 
First, the  precoding matrix at the BS and reflection coefficient at the IR are jointly optimized to  maximize the downlink sum-rate towards multiple users.   
Next,  to address the uncertainty of mmWave channels and maintain LOS links in a real-time manner, a distributional reinforcement learning approach \cite{dabney2018distributional}, based on quantile regression optimization, is introduced to learn the mmWave communication environment, such that the location of the UAV-IR is optimized  to maximize the communication sum-rate over a long-term horizon.  
Simulation results show that the proposed learning-based deployment of the UAV-IR yields a significant advantage, compared to a non-learning UAV-IR, a static IR and a direct transmission schemes, in terms of the average data rate and the achievable downlink LOS probability.

The rest of this paper is organized as follows. Section \ref{sysModel_proFormulation} presents the system model and problem formulation. The optimal deployment of the UAV-IR for  multi-user downlink transmissions is proposed in Section \ref{solution}. Simulation results are shown in Section \ref{simulation}. Conclusions are drawn in Section \ref{conclusion}.

\section{System Model and Problem Formulation}\label{sysModel_proFormulation}
 
\begin{figure}[!t]
	\begin{center}
		\vspace{-0.2cm}
		\includegraphics[width=7cm]{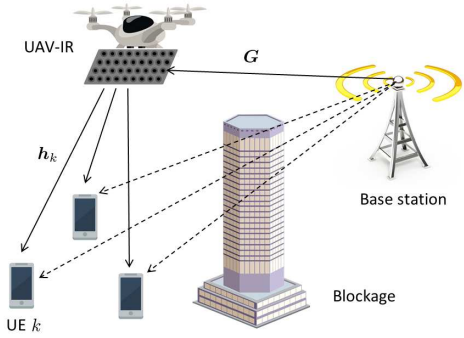}
		\vspace{-0.6cm}
		\caption{\label{systemmodel}\small A UAV-IR establishes LOS links between the BS and UEs for efficient downlink transmissions over mmWave frequencies. }  
	\end{center}\vspace{-0.7cm}  
\end{figure}

Consider a wireless BS serving a set $\mathcal{K}$ of $K$  cellular user equipment (UE) over the downlink via mmWave frequencies.   
The BS has  $M$  transmit antennas, and each downlink UE has a single antenna.  
As shown in Fig. \ref{systemmodel}, we assume that the direct transmission channel between the BS and each UE is blocked, and the received signal via the direct link is negligible.    
In order to bypass the obstacle and improve the received power at each UE,  a UAV-IR with $N$ reflective elements is deployed to assist  downlink transmissions towards NLOS UEs.      
By leveraging the mobility of the UAV, the UAV-IR can potentially replace each direct NLOS link with two connected LOS links, by adjusting its position and reflecting mmWave signals from the BS towards each UE.        
Since an IR is a passive device, it cannot acquire the channel state information (CSI) or process received signals. 
To enable information exchange, an active antenna is embedded onto the UAV to receive a control signal from the BS and feedback information from served UEs.

\subsection{Communications Capacity and Power Cost}  

Consider a multi-user multiple-input-single-output downlink communications, in which the BS serves downlink UEs via a common mmWave band. 
The BS-IR channel is denoted as $\boldsymbol{G} \in \mathbb{C}^{N\times M}$, and the IR-UE link for each UE $k$ is given as  $\boldsymbol{h}_k \in \mathbb{C}^{1\times N}$.       
Let  $\mathcal{N}$ be the set of $N$ IR components. 
For each component $n \in \mathcal{N}$,  the phase shift is denoted by $\phi_n \in [0,2\pi) $ and the amplitude attenuation is $\beta_n  \in [0,1]$.  
Consequently, the IR's reflection coefficient is   $\Phi = \text{diag} (\beta_1  e^{j\phi_1},\cdots, \beta_N e^{j\phi_N} )$.  
In order to provide downlink communications to multiple UEs, the BS precodes the transmit signal as an $M\times 1$ vector $\boldsymbol{a} = \sum_{k=1}^{K} \boldsymbol{w}_k s_k$, where  $\boldsymbol{w}_k \in \mathbb{C}^{M\times 1}$ is the precoding vector,  and $ s_k$ is the unit-power information symbol for UE $k$. 
The  received signal at UE $k$ will thus be:    
$	y_k =  \boldsymbol{h}_k \Phi \boldsymbol{G} \boldsymbol{a} + u_k$, 
where  $u_k \sim \mathcal{CN}(0,\sigma^2)$ is the receiver noise at UE $k$.  
In order to separate the mmWave propagation environment  with the IR's reflection, we rewrite the received signal at UE $k$ equivalently in the following form:\vspace{-0.2cm}
\begin{equation}\vspace{-0.2cm}
  	y_k =  \boldsymbol{\theta}\boldsymbol{D}_k \boldsymbol{a} + u_k, 
\end{equation}
where  $\boldsymbol{\theta} = [\beta_1e^{j\phi_1},\cdots, \beta_N e^{j\phi_N}] \in \mathbb{C}^{1 \times N}$ is a vector of the UAV-IR reflection coefficients and $\boldsymbol{D}_k = \text{diag} ( \boldsymbol{h_k}) \boldsymbol{G} \in \mathbb{C}^{N \times M} $ is the CSI of the connected BS-IR-UE link towards UE $k$ without any phase shift.  
The channel measurement approach for IR-aided cellular communications has been investigated in \cite{zhang2020millimeter}. 
Here, we assume that the CSI $\boldsymbol{D}_k(\boldsymbol{x})$ of each BS-IR-UE link only depends on the location $\boldsymbol{x}$ of the UAV-IR, and the channel matrix $\boldsymbol{D}_k(\boldsymbol{x})$  is known to both BS and UAV-IR, as long as  $\boldsymbol{x}$  is given. 
Therefore, the signal-to-interference-and-noise-ratio (SINR) of downlink communications from the BS, reflected by the IR, to each UE $k$ is   
\begin{equation}\label{sinr} 
\begin{aligned}
\eta_k ( \boldsymbol{W}, \boldsymbol{\theta}, \boldsymbol{x}) 
&= 	
\frac{  |\boldsymbol{\theta} \boldsymbol{D}_k(\boldsymbol{x}) \boldsymbol{w}_k |^2 }{  \sum_{i\ne k, i\in \mathcal{K}} |\boldsymbol{\theta} \boldsymbol{D}_k(\boldsymbol{x}) \boldsymbol{w}_i  |^2  + \sigma^2},  
\end{aligned} 
\end{equation}  
where
$\boldsymbol{W} = [\boldsymbol{w}_1,\cdots,\boldsymbol{w}_K] \in \mathbb{C}^{M \times K} $ is a  precoding matrix at the BS.  
Consequently, the total achievable rate that the IR-assisted communication can provide to all UEs is  
\begin{align} 
C (\boldsymbol{W}, \boldsymbol{\theta}, \boldsymbol{x}) = \sum_{k=1}^{K} b\log_2 ( 1+  \eta_k( \boldsymbol{W},\boldsymbol{\theta}, \boldsymbol{x})),   
\end{align}  
where $b$ is the downlink bandwidth.

In order to maintain LOS links with both the BS and UEs, the UAV-IR needs to frequently adjust its location, based on the real-time CSI.   
Thus, the power cost of a UAV-IR includes the UAV's hovering power $p_h$, mobility power $p_m$, and the adjustment power $p_r$ for the IR's reflection coefficients.     
In order to facilitate beamforming transmissions at the BS and ensure a reliable reflection at the IR,  we assume that downlink communication only happens when the UAV-IR has a fixed location.   
Let  $v$ be the speed of the UAV-IR, and $\mathbbm{1}_{v =0}$ is an indicator function which equals to one when the UAV is in the hovering state with a zero speed. 
Therefore, the  power cost of the UAV-IR for providing downlink reflection service is  
$	p(v ) = \mathbbm{1}_{v =0} \cdot ( p_h + p_r ) + (1-\mathbbm{1}_{v =0}) \cdot  p_m$.  	 

\subsection{Problem Formulation} \label{problemFormulation}

\begin{figure}[!t]
	\begin{center}
		\vspace{-0.8cm}
		\includegraphics[width=7.8cm]{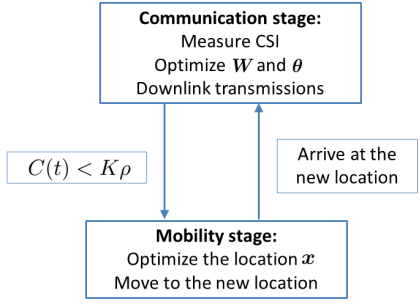}
		\vspace{-0.8cm}
		\caption{\label{process}\small The two-stage deployment process of the UAV-IR. }
	\end{center}\vspace{-0.7cm}  
\end{figure}

As shown in Fig. \ref{process}, in order to provide efficient and reliable  mmWave communications to downlink UEs, a dynamic deployment for the UAV-IR is considered, where the deployment process is divided into  two sequential and alternating stages: communication  and movement.      
In the communication stage, the UAV-IR stays at a fixed location while reflecting mmWave signals towards downlink UEs. 
However, once the average downlink rate is lower than a threshold $\rho$, blockage occurs in the downlink channels for most UEs.
In this case, the current communication stage ends, and the UAV-IR moves to a new position, so as to establish LOS links for efficient downlink communications.    
Here, we assume that the CSI $\boldsymbol{D}_k$ is constant within each coherence time slot $\Delta T$.  
Given that the duration of a communication stage depends on the real-time CSI  and the length of a movement stage is determined by the movement distance and the UAV's speed,  both stages can last for several coherence time slots.

Our goal is to jointly optimize the precoding matrix at the BS,  the reflection coefficient of the IR, and the location of the UAV, such that, before the UAV's onboard energy is exhausted, the total achievable data transmissions that the UAV-IR  provides to downlink UEs can be maximized, i.e.:   
\begin{subequations}\label{equsOpt} \vspace{-0.24cm}
\begin{align}
	\max_{\{ \boldsymbol{W}(t),\boldsymbol{\theta}(t),\boldsymbol{x}(t)\}_{\forall t}} \quad &  \sum_{t=1}^{T}   \mathbbm{1}_{v(t)=0} \cdot C \left(\boldsymbol{W}(t),\boldsymbol{\theta}(t),\boldsymbol{x}(t)\right) \cdot \Delta T  \label{equOpt}\\ 
	\textrm{s. t.} \quad  
	& \sum_{k=1}^{K}\|\boldsymbol{w}_k(t)\|^2 \le P_\textrm{max},  \label{consBSPower}\\ 
	&|\theta_n(t)| \le 1,  \quad \forall n \in \mathcal{N},  \label{consIR} \\
	& \sum_{t=1}^{T} p(v(t)) \cdot \Delta T \le E ,  \label{consUAVenergy} 
\end{align}
\end{subequations}
where the random integer $T \in \mathbb{N}^{+}$ denotes the final time slot of the UAV's service, $v(t) = \frac{\|\boldsymbol{x}(t-1)-\boldsymbol{x}(t)\|}{\Delta T}$ is the average speed of the UAV during the coherence time $t$, 
$ P_\textrm{max}$ is the maximal transmit power at the BS, 
and $E$ is the initial onboard energy of the UAV.   
Therefore, the objective function (\ref{equOpt}) is the summation of downlink data transmissions that the UAV-IR provides to all UEs before the end of its service, (\ref{consBSPower}) is the power limitation at the BS, (\ref{consIR}) is the reflection constraints at the IR, and (\ref{consUAVenergy}) is the energy constraint of the UAV.

The optimization problem in (\ref{equsOpt}) is  challenging to solve for two reasons. 
First, during each coherence time, the objective function (\ref{equOpt}) is non-convex with respect to the optimal variables $\boldsymbol{W}$, $\boldsymbol{\theta}$ and $\boldsymbol{x}$.  
Second, the relationship between the location $\boldsymbol{x}$ of the UAV-IR with the CSI $\boldsymbol{D}_k(\boldsymbol{x})$ is not explicitly known.   
Meanwhile, it is impractical to apply a sweeping search by moving the UAV-IR to all possible locations and measuring the real-time CSI $\boldsymbol{D}_k$ for each UE $k$.  
In order to address these challenges, in Section \ref{solution1}, the beamforming matrix $\boldsymbol{W}$ and reflection coefficient $\boldsymbol{\theta}$ will be optimized, given a fixed location and known CSI. 
Next, the location optimization of the UAV-IR will be addressed using a learning-based approach to model the mmWave communication environment.

\section{Optimal Deployment of UAV-IR} \label{solution}

In this section, the precoding matrix at the BS, the reflection coefficients at the IR, and the location of the UAV-IR will be jointly optimized in a real-time manner to maximize the downlink transmission capacity. 
In particular, a  distributional reinforcement learning (DRL) framework is proposed  to model the downlink CSI during each communication stage,  based on the UEs' feedback.

\subsection{Optimal precoding and reflection coefficients}  \label{solution1}

First, we focus on the communication stage, where the UAV-IR has a fixed location $\boldsymbol{x}$ and the downlink CSI $\boldsymbol{D}_k$ for each UE $k$ is known.   
Then,  for each coherence time,  (\ref{equsOpt}) is reduced to optimize the precoding matrix and the reflection coefficients, so as to maximize the downlink sum-rate, i.e.:  
\begin{subequations}\label{equsOptsub}
	\begin{align}
	\max_{  \boldsymbol{W} ,\boldsymbol{\theta}  } \quad &   C \left(\boldsymbol{W} ,\boldsymbol{\theta}  \right)  \label{equOptsub}\\
	\textrm{s. t.} \quad  
	& \sum_{k=1}^{K}\|\boldsymbol{w}_k \|^2 \le P_\textrm{max},  \label{consBSPowersub}\\ 
	& |\theta_n | \le 1,  \quad \forall n \in \mathcal{N}.  \label{consIRsub} 
	\end{align}
\end{subequations}
To solve this non-convex problem, we apply the Lagrangian dual transform method \cite{shen2018fractional}, introduce an auxiliary variable $\boldsymbol{\alpha} \in \mathbb{R}^{K\times 1}$, and equivalently rewrite the objective function (\ref{equOptsub}) into the following form:
\begin{equation}\label{sumrateAlpha}
	C_{\alpha} (\boldsymbol{W} ,\boldsymbol{\theta}, \boldsymbol{\alpha}) = b\sum_{k=1}^K \left(\log_2 (1 + \alpha_k)   -\alpha_k + \frac{(1+\alpha_k) \eta_k}{1+\eta_k}  \right),
\end{equation} 
By holding $\boldsymbol{W}$ and $\boldsymbol{\theta}$  fixed and setting $\frac{\partial  C_{\alpha} }{\partial  \alpha_k} = 0$, we have the optimal value of $\alpha_k$ as $\alpha^o_k = \eta_k$. Then, for  a fixed $\boldsymbol{\alpha}$, the optimization problem of $\boldsymbol{W}$ and $\boldsymbol{\theta}$ is reduced to
\begin{subequations}\label{equsOptsub2}\vspace{-0.2cm}
	\begin{align}
	\max_{  \boldsymbol{W} ,\boldsymbol{\theta} } \quad &   \sum_{k=1}^K  \frac{\hat{\alpha}_k \eta_k}{1+\eta_k}   \label{equOptsub2}\\
	\textrm{s. t.} \quad  & (\text{\ref{consBSPowersub}}), (\text{\ref{consIRsub}}), 
	\end{align}
\end{subequations}
where $\hat{\alpha}_k = b (1+\alpha_k)$. 
Given that (\ref{equsOptsub2}) is a multiple-ratio fractional programming problem, we can fix the value of $\boldsymbol{W}$ and $\boldsymbol{\theta}$ alternatively, and solve the optimization problem via an iterative approach, detailed as follows.  

\subsubsection{Optimal precoding matrix}\label{optPrecoding}
For a fixed $\boldsymbol{\theta}$, the optimal precoding problem becomes
\begin{subequations}\label{equsOptsub3}
	\begin{align}
	\max_{  \boldsymbol{W}   } \quad &  f(\boldsymbol{W}) = \sum_{k=1}^K \hat{\alpha}_k  \frac{  |\boldsymbol{\theta} \boldsymbol{D}_k  \boldsymbol{w}_k |^2 }{  \sum_{i=1 }^{K} |\boldsymbol{\theta} \boldsymbol{D}_k  \boldsymbol{w}_i  |^2  + \sigma^2}  \label{equOptsub3}\\
	\textrm{s. t.} \quad  & \sum_{k=1}^{K}\|\boldsymbol{w}_k \|^2 \le P_\textrm{max}. 
	\end{align}
\end{subequations} 	
The multiple-ratio fractional programming function in (\ref{equOptsub3}) is equivalent to $f_{\lambda}(\boldsymbol{W},\boldsymbol{\lambda}) = $  
$- \sum_{k=1}^K |\lambda_k|^2  (\sum_{i=1}^K | \boldsymbol{\theta} \boldsymbol{D}_k  \boldsymbol{w}_i |^2 + \sigma^2) $ $+$ $ \sum_{k=1}^K   2 \sqrt{\hat{\alpha}_k} \text{Re} \{ \lambda_k  \boldsymbol{\theta} \boldsymbol{D}_k  \boldsymbol{w}_k  \}$,   
where  $\boldsymbol{\lambda} \in \mathbb{R}^{K \times 1}$ is an auxiliary vector \cite[Theorem 2]{shen2018fractional}. 
Since $f_{\lambda}$ is a convex function with respect to both $\boldsymbol{w}_k$ and ${\lambda}_k$, $\forall k$, an iterative approach can be applied to optimize  $\boldsymbol{W}$ and $\boldsymbol{\lambda}$ alternatively. 
First, we fix $\boldsymbol{W}$ and set $\frac{\partial f_{\lambda} }{\partial \lambda_k}=0$. Then, the optimal value of $\lambda_k$ is 
\begin{equation}\label{zerolambda}
	\lambda^o_k = \frac{\sqrt{\hat{\alpha}_k}  \boldsymbol{\theta} \boldsymbol{D}_k  \boldsymbol{w}_k}{\sum_{i=1}^K | \boldsymbol{\theta} \boldsymbol{D}_k  \boldsymbol{w}_i |^2 + \sigma^2 }.
\end{equation} 
Then, by fixing $\boldsymbol{\lambda}$, the optimal $\boldsymbol{w}_k$ can be given by 
\begin{equation} \label{zerow} \vspace{-0.1cm}
	\boldsymbol{w}_k^o = \sqrt{\hat{\alpha}_k} \lambda_k \boldsymbol{\theta} \boldsymbol{D}_k \left( \kappa_o \boldsymbol{I}_M + \sum_{i=1}^K |\lambda_i|^2 (\boldsymbol{\theta} \boldsymbol{D}_i )(\boldsymbol{\theta} \boldsymbol{D}_i )^H    \right)^{-1}   ,
\end{equation}
where $\kappa_o \ge 0$ is the minimum value such that $\sum_{k=1}^K\| \boldsymbol{w}_k^o \|^2 \le P_{\text{max}}$ holds. 
By alternating between (\ref{zerolambda}) and (\ref{zerow}),  the precoding matrix will eventually converge to the unique and optimal value $\boldsymbol{W}^{*}$,  with a computational complexity of $\mathcal{O}(M^4)$. 

\subsubsection{Optimal reflection coefficients}
Next, we fix the value of $\boldsymbol{W}$ and optimize the reflection coefficient $\boldsymbol{\theta}$ in (\ref{equsOptsub2}), i.e.:    
\begin{subequations}\label{equsOptsub4} \vspace{-0.1cm}
	\begin{align}
	\max_{  \boldsymbol{\theta}   } \quad &  f(\boldsymbol{\theta}) = \sum_{k=1}^K \hat{\alpha}_k  \frac{  |\boldsymbol{\theta} \boldsymbol{D}_k  \boldsymbol{w}_k |^2 }{  \sum_{i=1 }^{K} |\boldsymbol{\theta} \boldsymbol{D}_k  \boldsymbol{w}_i  |^2  + \sigma^2}  \label{equOptsub4}\\
	\textrm{s. t.} \quad  &  |\theta_n | \le 1,  \quad \forall n \in \mathcal{N}.  
	\end{align}
\end{subequations}
Similarly, an auxiliary vector $\boldsymbol{\delta}$ is introduced, so that (\ref{equOptsub4}) equivalently becomes  
$f_{\delta} (\boldsymbol{\theta},\boldsymbol{\delta}) =  \sum_{k=1}^K   2 \sqrt{\hat{\alpha}_k} \text{Re} \{ \delta_k  \boldsymbol{\theta} \boldsymbol{D}_k  \boldsymbol{w}_k  \} $ $ - \sum_{k=1}^K |\delta_k|^2  (\sum_{i=1}^K | \boldsymbol{\theta} \boldsymbol{D}_k  \boldsymbol{w}_i |^2 + \sigma^2)$, which is convex with respect to both $\boldsymbol{\theta}$ and $\boldsymbol{\delta}$.   
Therefore, the unique and optimal value of the reflection coefficients $\boldsymbol{\theta}^*$ can be obtained via a similar alternative approach as the precoding  optimization. 
First, the optimal $\delta_k$ for a given $\boldsymbol{\theta}$ is 
\begin{equation}\label{zerodelta}
	\delta_k^o = \frac{\sqrt{\hat{\alpha}_k}  \boldsymbol{\theta} \boldsymbol{D}_k  \boldsymbol{w}_k}{\sum_{i=1}^K | \boldsymbol{\theta} \boldsymbol{D}_k  \boldsymbol{w}_i |^2 + \sigma^2 }. 
\end{equation}
Then, the optimization of the reflection coefficient for a fixed $\delta_k$ becomes
\begin{subequations}\label{equsOptsub5}\vspace{-0.2cm}
	\begin{align}
	\max_{  \boldsymbol{\theta}   } \quad &  f(\boldsymbol{\theta}) =  -\boldsymbol{\theta} \boldsymbol{U} \boldsymbol{\theta}^H + 2\text{Re}\{ \boldsymbol{\theta} \boldsymbol{v} \} -C  \label{equOptsub5}\\
	\textrm{s. t.} \quad  &  |\theta_n | \le 1,  \quad \forall n \in \mathcal{N},   \label{cons}
	\end{align}
\end{subequations}
where  $C = \sum_{k=1}^K |\delta_k|^2\sigma^2$,  $\boldsymbol{v} = \sum_{k=1}^K {\delta_k}^H \boldsymbol{D}_k \boldsymbol{w}_k$, and $\boldsymbol{U} = \sum_{k=1}^K |\delta_k|^2 \sum_{i\ne k}  \boldsymbol{D}_k  \boldsymbol{w}_i ( \boldsymbol{D}_k  \boldsymbol{w}_i)^H $. 
Given that $\boldsymbol{U}$ is a positive-definite matrix,   $ f(\boldsymbol{\theta})$ is quadratic concave with respect to $\boldsymbol{\theta}$. 
Meanwhile, the constraint (\ref{cons}) is a convex set.
Thus, (\ref{equsOptsub5}) is solvable using a Lagrange dual decomposition \cite{guo2019weighted} with a computational complexity of $\mathcal{O}(N^6)$.      

Therefore, given the CSI $\boldsymbol{D}_k$ of each BS-IR-UE link,  the precoding matrix $\boldsymbol{W}^*$ and reflection coefficient $\boldsymbol{\theta}^{*}$ can be optimally  and uniquely  determined.  
Next,  the optimal location  of the UAV-IR will be studied to guarantee a LOS channel for each BS-IR-UE link.

\subsection{Optimal location of the UAV-IR}\label{solution2}
During the communication stage, at the end of each coherence time $t$, the UAV-IR can get feedback from each UE about the downlink transmission performance.  
Whenever the average rate per UE is smaller than $\rho$, i,e, $C(t) < K\rho$,  blockage occurs in the downlink channels for most UEs. 
In this case, the UAV-IR needs to move and rebuild LOS links.  
The optimization problem of the UAV-IR's location is given as
\begin{subequations}\label{equsOptloc} \vspace{-0.1cm}
	\begin{align}
	\max_{\{ \boldsymbol{x}(t)\}_{\forall t}} \quad &  \sum_{t=1}^{T}   \mathbbm{1}_{v(t)=0} \cdot C \left(\boldsymbol{x}(t)\right) \cdot \Delta T  \label{equOptloc}\\ 
	\textrm{s. t.} \quad   
	& \sum_{t=1}^{T} p(v(t)) \cdot \Delta T \le E .  \label{consUAVenergyloc} 
	\end{align}
\end{subequations}

To determine the optimal location for each communication stage,  a DRL framework is designed to learn the dynamic communication environment and model the relationship between the UAV's location and  downlink CSI.  
In our DRL framework, the downlink CSI $\{ \boldsymbol{D}_k\}_{\forall k}$ is the communication \emph{environment}, 
the UAV-IR is the \emph{agent} that takes \emph{action} $\Delta \boldsymbol{x}$ to change its location from $\boldsymbol{x}$ to  $\boldsymbol{x} +\Delta \boldsymbol{x}$,   
and the communication \emph{state} $\boldsymbol{e}$ is a vector of the received signal power at each UE, where $\boldsymbol{e} = [|y_1|^2, \cdots, |y_K|^2]$. 
At the end of each coherence time, the UAV-IR receives \emph{reward}  $r(\boldsymbol{x}) = \mathbbm{1}_{v=0} C(\boldsymbol{x}) \Delta T$, which is the total received data in the downlink transmission.  
Due to the small‐scale fading of mmWave channels, the downlink CSI $\boldsymbol{D}_k(\boldsymbol{x})$ may vary between different time slots, even for a fixed location $\boldsymbol{x}$ of the UAV-IR.    
Thus, it is more suitable to consider the  reward $r$ as a random variable  with respect to $\boldsymbol{x}$, rather than a determined value.  
Meanwhile,  $P(\boldsymbol{e}^{'}|\boldsymbol{e},\Delta \boldsymbol{x})$ is the state transition probability  from $\boldsymbol{e}$  to $\boldsymbol{e}^{'}$ after taking action $\Delta \boldsymbol{x}$. 

To properly capture the relationship between the UAV's movement and downlink transmission performance, first, a \emph{policy} $\pi(\Delta \boldsymbol{x} | \boldsymbol{e})$ is introduced to define the probability that the UAV-IR will move by $ \Delta \boldsymbol{x}$,  under a current state $\boldsymbol{e}$. 
Meanwhile, to quantify the potential of each action $\Delta \boldsymbol{x}$ for improving the downlink rate under a state $\boldsymbol{e}$, we define the return function of each state-action pair for any time slot $t$ as  
\begin{equation}\label{Qfunction}
\begin{aligned}
Z^{\pi}(\boldsymbol{e}_t,\Delta\boldsymbol{x}_t) =      \sum_{i=t}^{T} \gamma^{i-t} r(\boldsymbol{e}_i,\Delta\boldsymbol{x}_i) , 
\end{aligned}
\end{equation}
where $\Delta \boldsymbol{x}_i \sim \pi(\cdot|\boldsymbol{e}_i)$, $\boldsymbol{e}_{i+1} \sim P(\cdot|\boldsymbol{e}_i,\Delta \boldsymbol{x}_i)$ and $\boldsymbol{x}_{i+1} = \boldsymbol{x}_i + \Delta \boldsymbol{x}_i$. 
Here, $\gamma \in (0,1)$  discounts the future rewards in the current estimation for each state-action pair.  
If $\gamma \rightarrow 1$, the return function $Z^{\pi}$ will approximate  (\ref{equOptloc}).  
Thus, the return function (\ref{Qfunction}) defines a cumulative discounted reward that the UAV-IR can achieve by reflecting mmWave signals at location $\boldsymbol{x}_t + \Delta\boldsymbol{x}_t$ for the next communication stage.  
Meanwhile, given that $r$ is a random variable, it is necessary to model a distribution function of (\ref{Qfunction}) to identify the return value for each state-action pair.   
Once the return distribution  is known, the optimal policy $\pi$ that maximizes the expectation of the cumulative rewards can be defined by   
$\Delta \boldsymbol{x}^{*} = \arg \max_{\Delta \boldsymbol{x}}  \mathbb{E}(Z^{\pi}(\boldsymbol{e}_t,\Delta\boldsymbol{x}_t ))$.   
Thus, the optimal location of the UAV-IR for the next time slot will be $\boldsymbol{x}_{t+1} =\boldsymbol{x}_{t} + \Delta \boldsymbol{x}^{*} $.

In order to model the return distribution for each state-action pair, a quantile regression (QR) method \cite{dabney2018distributional} is applied.
A $Q$-quantile model $Z_Q$ approximates the target distribution $Z^{\pi}$, using a discrete function with variable locations of $Q$ supports and fixed quantile  of  $\frac{1}{Q}$ probabilities \cite{zhang2020millimeter}, for a fixed integer $Q \in \mathbb{N}^{+}$. 
Mathematically, a $Q$-quantile model is denoted by $Z_Q(\boldsymbol{e},\Delta\boldsymbol{x} ) = [z_1(\boldsymbol{e},\Delta\boldsymbol{x} ), \cdots , z_Q(\boldsymbol{e},\Delta\boldsymbol{x} )] $, with a cumulative probability $F_{Z_Q}(z_q) = \frac{q}{Q}$ for $q = 1,\cdots, Q$. 
The objective is to find the optimal location for each support, such that the ``distance'' between  the target distribution $Z^{\pi}$ and the $Q$-quantile model $Z_Q$ can be minimized.  %
However, given that the actual return distribution $Z^{\pi}$  is not explicitly known, an empirical distribution $Z$ will be formed, based on the UEs' feedback during each time slot, and $Z$ is used as the target distribution to model the return approximation  $Z_Q$.  
To quantify the ``distance'' between two distribution functions, the quantile regression loss is defined as \cite{dabney2018distributional},
\begin{equation}
	\mathcal{L}_Z(Z_Q) = \sum_{q=1}^Q \mathbb{E}_Z \left[|\omega_q - \mathbbm{1}_{z<z_q}| \cdot (z - z_q)^2 \right],
\end{equation}
where $\omega_q = \frac{2q-1}{2Q}$, $|\omega_q - \mathbbm{1}_{z<z_q}| $ is the weight of regression loss penalty, and $(z - z_q)^2$ is the square of approximation error.  
Thus, the  problem of the return distribution modeling becomes to minimize the quantile regression loss, i.e.,
\begin{equation}\label{1WminProj} 
\min_{ z_1, \cdots, z_Q } 	\mathcal{L}_Z(Z_Q).  
\end{equation}
Since the objective function in (\ref{1WminProj}) is convex with respect to $Z_Q$, the minimizer $\{ z^{*}_q \}_{q=1,\cdots,Q}$ can be found by conventional  gradient-descent approaches  with a computational complexity of $\mathcal{O}(Q^2)$. 
As a result, for each state-action pair, its return distribution $Z_Q(\boldsymbol{e},\Delta \boldsymbol{x}) $ can be approximated by a  Q-quantile $\{ z^{*}_1(\boldsymbol{e},\Delta \boldsymbol{x}) , \cdots, z^{*}_Q(\boldsymbol{e},\Delta \boldsymbol{x})  \}$ via (\ref{1WminProj}).

Consequently, in the location optimization problem of the UAV-IR, after observing a communication state $\boldsymbol{e}_t$, the UAV-IR can estimate the expected return value for each action $\Delta \boldsymbol{x}$, by computing the marginal distribution of  $Z_Q(\boldsymbol{e},\Delta \boldsymbol{x}) $, and choose the optimal location $\boldsymbol{x}_{t+1} = \boldsymbol{x}_t + \Delta \boldsymbol{x}^{*} $ that maximize the summation of future downlink  transmissions via
\begin{equation}
\begin{aligned} 
\Delta \boldsymbol{x}^{*} = \arg \max_{\Delta \boldsymbol{x}}\mathbb{E} \left[  Z_Q(\boldsymbol{e}_t,\Delta \boldsymbol{x})  \right] 
=   \arg \max_{\Delta \boldsymbol{x}} \frac{1}{Q} \sum_{q=1}^Q z_q(\boldsymbol{e}_t,\Delta \boldsymbol{x}). 
\end{aligned}
\end{equation}

After the UAV-IR arrives at the new location $\boldsymbol{x}_{t+1}$ and provides downlink service to UEs, a new state $\boldsymbol{e}_{t+1}$ and reward $r_{t+1}$ will be updated  at the end of the $l+1$ time slot. 
Given the downlink transmission,  the empirical distribution can be updated via a Q-learning approach, where 
$z_i(\boldsymbol{e}_t,\Delta \boldsymbol{x}_t) \leftarrow r_{t+1} + \gamma z_i^t(\boldsymbol{e}_{t+1},\Delta \boldsymbol{x}_{t+1}), \forall i = 1,\cdots,Q$. 
As a result, the return distribution $Z^t_Q$ is updated by minimize the distance from the target distribution $Z$, based on (\ref{1WminProj}). 
The training and update algorithm of the DRL model for the real-time optimal deployment of the UAV-IR  is summarized in Algorithm \ref{algo}. 
The convergence property of this iterative algorithm has been proved  in \cite[Theorem 1]{zhang2020millimeter}.

\begin{algorithm}[h] \small   
\caption{DRL-optimized deployment for the UAV-IR } \label{algo}
\begin{algorithmic}
	\State \textbf{Initialize} the precoding matrix $\boldsymbol{W}_0$, the reflection coefficient $\boldsymbol{\theta}_0$, the location $\boldsymbol{x}_0$, the onboard energy $E_0$, and the DRL model  function $Z^0_{Q}(\boldsymbol{e}, \Delta \boldsymbol{x})$ for each state-action pair. \\
	\textbf{For} $t = 1, \cdots, T$:\\
	\quad A. If $v_t>0$, the UAV-IR continues its movement; \\
	\quad B .If $v_t=0$, the UAV-IR is in the communication stage;\\ 
	\quad \textbf{Repeat}:\\
	\quad \quad B1. Update the auxiliary variable $\alpha_k = \eta_k$; \\
	\quad \quad B2. Optimize $\boldsymbol{W}_t^{*}$ by alternating between (\ref{zerolambda}) and (\ref{zerow}); \\ 
	\quad \quad B3. Optimize $\boldsymbol{\theta}_t^{*}$ , by alternating between (\ref{zerodelta}) and (\ref{equsOptsub5}); \\
	\quad \textbf{Until} the value of $C_{\alpha}$ in (\ref{sumrateAlpha}) converges;  \\
	\quad \quad B4. Receive the reward $r_t$ and the  state $\boldsymbol{e}_t$.  If $r_t \ge K\rho \Delta T$,  \\
	\quad  \quad \quad  $\boldsymbol{x}_{t+1} = \boldsymbol{x}_t$;   Otherwise, the UAV-IR moves by \\
	\quad \quad \quad \quad $\Delta \boldsymbol{x}^{*} = \arg \max_{ \Delta \boldsymbol{x}} \frac{1}{Q} \sum_{q=1}^Q z^{t-1}_q (\boldsymbol{e}_t,\Delta \boldsymbol{x}) $.   \\
	\quad \quad B5. Update the empirical distribution $Z$ via \\  
	\quad \quad \quad  $z_i(\boldsymbol{e}_{t-1},\Delta \boldsymbol{x}_{t-1})  \leftarrow r_t + \gamma z_i^{t-1}(\boldsymbol{e}_t,\Delta \boldsymbol{x}_t)$, $\forall i=1,\cdots,Q$. \\ 
	\quad \quad B6. Update the DRL model  $Z^t_Q$  via,   \\  
	\quad \quad \quad $ \arg \min_{ \{z_q\}_{\forall q} }  \sum_{q=1}^Q \sum_{i=1 }^Q |\frac{2q-1}{Q} - \mathbbm{1}_{z_i<z_q}|\cdot (z_i-z_q)^2$. \\ 
	\quad C. Update onboard energy via  
	$E_{t+1} = E_t -p(v_t)\Delta T$. \\	
	\textbf{Until} $E_t =0$. 
\end{algorithmic}
\end{algorithm}

\begin{figure*}[!t]
	\begin{center}\vspace{-0.15cm}
		\begin{subfigure}{.325\textwidth}
			\centering
			\includegraphics[width=6.4cm]{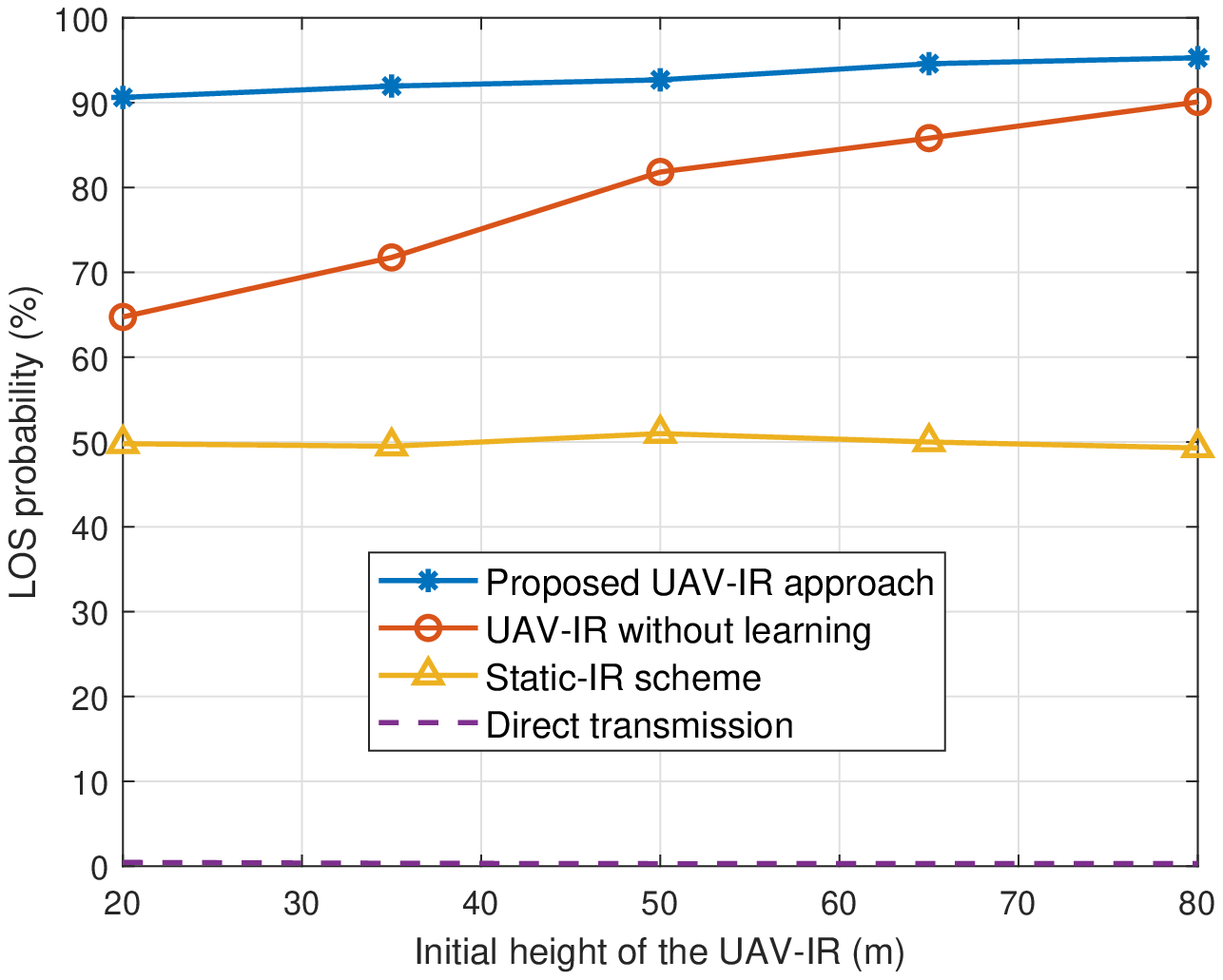}
			\caption{\label{probability} }
		\end{subfigure}
		\begin{subfigure}{.325\textwidth}
			\centering
			\includegraphics[width=6.4cm]{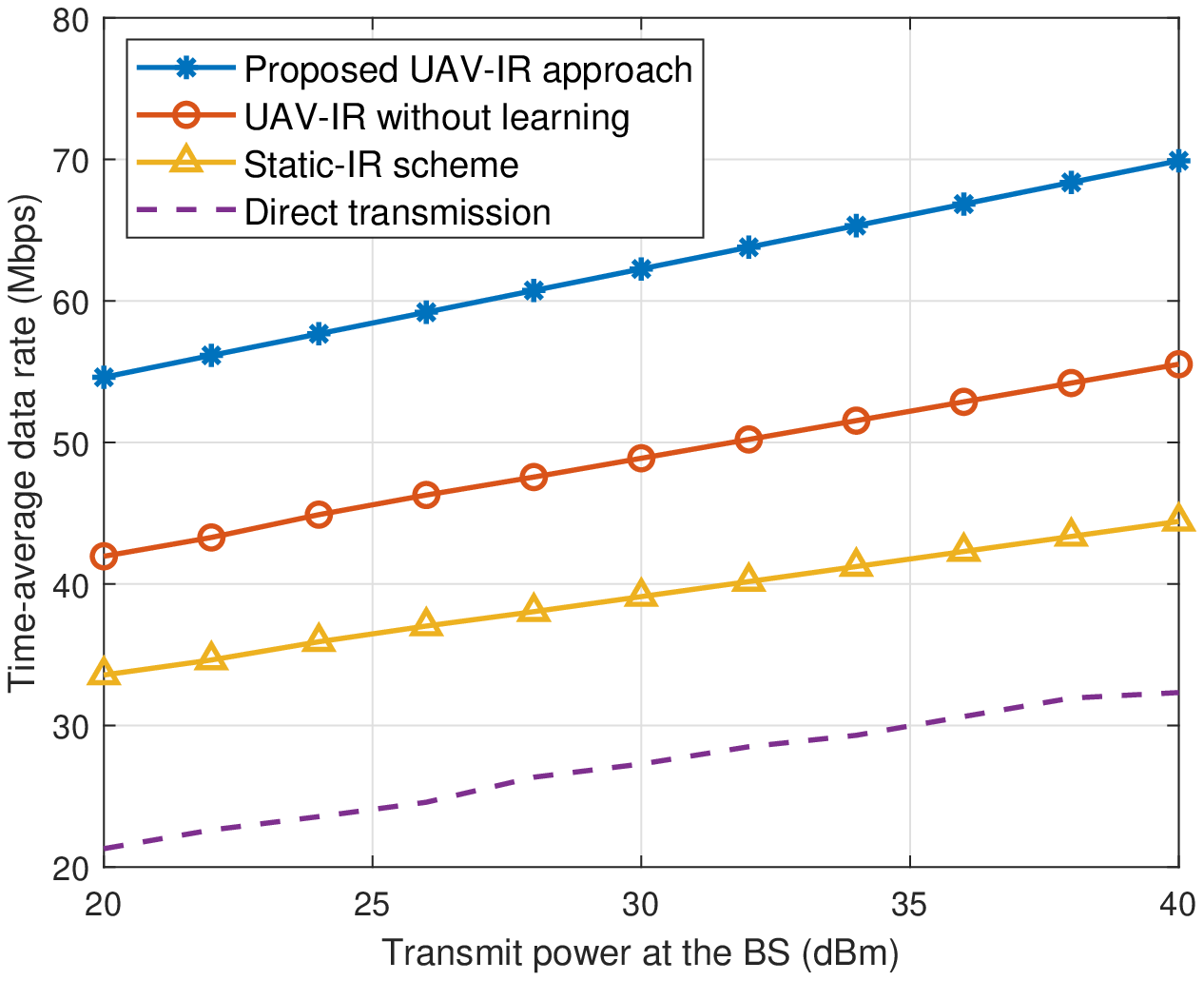}
			\caption{\label{power}  }
		\end{subfigure}
		\begin{subfigure}{.33\textwidth}
			\centering
			\includegraphics[width=6.5cm]{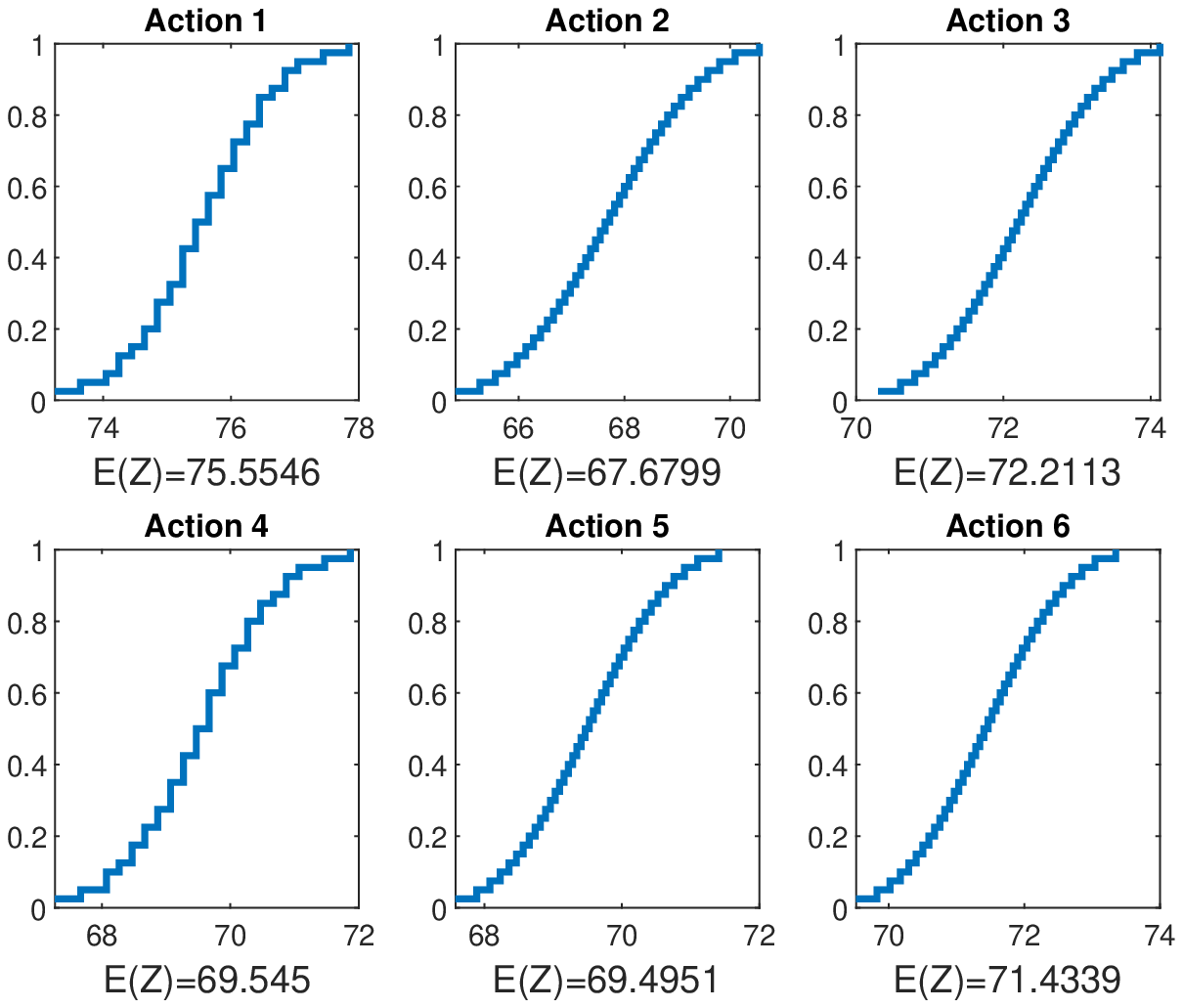}
			\caption{\label{distributions} }
		\end{subfigure}
		\vspace{-0.15cm}
		\caption{\small{\label{fig:both} (a) The achievable LOS probability of downlink mmWave channels. (b) The downlink rate increases, as the transmit power at the BS  increases. (c) The return distribution of each action under the worst-case communication state $\boldsymbol{e} = \boldsymbol{0}_K$. }
		}
	\end{center}
	\vspace{-0.4cm}
\end{figure*}
 
\section{Simulation Results and Analysis}\label{simulation}

For our simulations, we consider a uniform square array of antennas at both the BS and the UAV-IR with $M = 16$  and  $N = 16$, and the number of downlink UEs is $K =4$.  
The BS is located at $(0,0,20)$, each UE's location follows an i.i.d. two-dimensional Gaussian $\mathcal{N}((20,0),8\boldsymbol{I}_2)$ with zero height, and the mobility pattern of each UE follows a Markov decision process in \cite{zhang2019reflections}.  
A building located  at $(10,0,0)$ with a height of $18$ meters permanently blocks direct BS-UE links. 
The path loss and mmWave channel model  are based on  \cite{zhang2020millimeter}. 
For communication parameters, we set   $f = 30$ GHz,  $b=2$ MHz,  $P_{\text{max}}=40$ dBm, and  $\Delta T = 0.1$ second.  
The average speed of the UAV during the mobility stage is $v = 10$ m/s, the maximal onboard energy is $E_{\text{max}} = 20$ Wh, and the power cost are $p_m = 20$ W, $p_h = 16$ W, and $p_r = 0.16$ W. 
In the DRL model, we set $Q = 40$, and $\gamma = 0.9$.   
In order to have a finite state-action  space, we  discretize the communication state $\boldsymbol{e}$ to be a binary vector, where  if the received power of UE $k$ is smaller than a threshold $\tau$, $e_k = 0$; otherwise $e_k = 1$.  
The discrete action space is defined as:  $\Delta \boldsymbol{x}_1$ is ``ascend by one meter'', $\Delta \boldsymbol{x}_2$ is ``descend by one meter'', and $\Delta \boldsymbol{x}_3$ to $\Delta \boldsymbol{x}_{2+K}$ are ``move towards UE $k$ by one meter'' for each $k \in \mathcal{K}$.


In Fig. \ref{probability}, we first show the empirical probability of having a LOS downlink channel towards each UE. 
To evaluate the performance of the proposed DRL approach for the UAV-IR deployment, a direct transmission, a static IR, and a UAV-IR without learning are introduced as baselines.   
The static IR is placed at $(20,10,20)$ to bypass the building and establish LOS BS-IR-UE links. However, the bodies of human users may block mmWave channels. 
For the non-learning scheme, the UAV-IR moves towards a UE by one meter, every time downlink blockage occurs. 
As shown in Fig. \ref{probability}, the static IR scheme yields a LOS probability of around $50\%$. 
Due to the mobility of the UAV, the proposed UAV-IR scheme can maintain a LOS probability of over $90\%$, and the non-learning UAV-IR  results in a probability of over $60\%$.  
Moreover, as the altitude of the UAV-IR increases, the LOS downlink probability will naturally increase for both UAV-IR schemes.  
However, since the DRL-based deployment can estimate the potential of each action in improving the mmWave communication over a long term, the proposed  method always yields a higher LOS probability than the non-learning  scheme.


Fig. \ref{power} shows the time-average  data rate of mmWave  downlink communications, as the transmit power of the BS increases.  
When the transmit power increases from $20$ to $40$ dBm, 
the downlink data rates of all schemes become higher. 
First,  compared with the direct transmission scheme,  the proposed UAV-IR approach yields a performance gain of over two-folds in the downlink data rate, due to a higher LOS probability. 
Meanwhile, the DRL-enabled deployment of the UAV-IR improves the communication performance by over $25\%$ and $50\%$, compared to the non-learning UAV-IR and the static IR, respectively. 
For the non-learning scheme, its short-sighted  strategy causes a frequent movement of the UAV-IR,  thus yielding a lower downlink rate.


Fig. \ref{distributions} shows how to choose the optimal action under a worst-case  state $\boldsymbol{e} = \boldsymbol{0}_K$, where the received power at each UE is lower than the threshold. 
In this case, Fig. \ref{distributions} shows the  return distribution and the expected return value of each action. 
Since $\Delta \boldsymbol{x}_1$ yields the highest expected reward, the optimal action under the worst-case communication state $\boldsymbol{e} = \boldsymbol{0}_K$ is to increase the altitude of the UAV-IR by one meter. 

\section{Conclusion}\label{conclusion}

In this paper, we have proposed  a novel DRL-enabled approach to the deploy a UAV-IR for efficient downlink transmissions over mmWave frequencies towards multiple UEs.  
To maximize the downlink sum-rate, the optimal precoding matrix at the BS and reflection coefficient of the IR have been derived.  
In order to model the propagation environment of mmWave communications, the DRL method has been proposed  to optimize the location of the UAV-IR, so as to maximize the downlink communication capacity.  
Simulation results show that the proposed DRL-based deployment of the UAV-IR yields a significant advantage, compared to a non-learning UAV-IR, a static IR, and a direct transmission schemes, in terms of the average data rate and the achievable downlink LOS probability.  
Future research will focus on  multiple UAV-IRs in outdoor communication scenarios with mobile  users.

\bibliographystyle{IEEEtran}
\bibliography{references}

\end{document}